\documentclass[10pt,aps,prl,twocolumn,preprintnumbers,nofootinbib]{revtex4-2}
\usepackage{amsmath,amssymb,mathtools}
\usepackage{hyperref}
\usepackage{xspace}
\usepackage{graphicx}
\usepackage{grffile}
\usepackage{enumitem}
\usepackage{comment}
\newcommand{\sect}[1]{%
\bigskip\noindent%
{\bfseries\upshape\rmfamily\boldmath{#1}.}---%
\ignorespaces%
}

\newcommand{\der}{\mathrm{d}}

\newcommand{\Ccal}{\mathcal{C}}
\newcommand{\DC}{\Delta C}
\newcommand{\hx}{\hat{x}}
\newcommand{\hz}{\hat{z}}
\newcommand{\qb}{\bar{q}}
\newcommand{\qp}{q^{\prime}}
\newcommand{\qbp}{\bar{q}^{\prime}}

\newcommand{\MeV}{\mathrm{MeV}}

\newcommand\MSbar{\overline{\mathrm{MS}}}
\newcommand{\Lar}{\mathrm{L}}

\newcommand{\HVBM}{HVBM}

\begin{document}

\preprint{ZU-TH 19/24}

\title{Polarized semi-inclusive deep-inelastic scattering at NNLO in QCD}%

\author{Leonardo Bonino}
\affiliation{%
  Physik-Institut, Universit\"at Z\"urich, Winterthurerstrasse 190, 8057 Z\"urich, Switzerland}%
\author{Thomas Gehrmann}
 \affiliation{%
  Physik-Institut, Universit\"at Z\"urich, Winterthurerstrasse 190, 8057 Z\"urich, Switzerland}%
\author{Markus L\"ochner}
\affiliation{%
  Physik-Institut, Universit\"at Z\"urich, Winterthurerstrasse 190, 8057 Z\"urich, Switzerland}%
 \author{ Kay Sch\"onwald}
 \affiliation{%
  Physik-Institut, Universit\"at Z\"urich, Winterthurerstrasse 190, 8057 Z\"urich, Switzerland}%
\author{Giovanni Stagnitto}
\affiliation{%
Universit\`{a} degli Studi di Milano-Bicocca \& INFN, Piazza della Scienza 3, 20216 Milano, Italy}%


\begin{abstract}
  Semi-inclusive hadron production in longitudinally polarized deep-inelastic lepton-nucleon scattering is a powerful tool for resolving the quark flavor decomposition of the 
  proton's spin structure.  
We present the full next-to-next-to-leading order (NNLO) QCD corrections to the coefficient functions of polarized semi-inclusive deep-inelastic scattering (SIDIS) in analytical form, enabling the use of SIDIS measurements in precision studies of the proton spin structure. The numerical impact of these corrections is illustrated by a comparison with data of polarized single-inclusive hadron spectra from the DESY HERMES and CERN COMPASS experiments.
\end{abstract}

\maketitle

\section{Introduction}

The proton is a complex bound state of quarks and gluons. Its internal structure can be described in a 
probabilistic manner in the parton model, formulated in the framework of the theory of strong interactions (quantum chromodynamics, QCD). Owing to a large wealth of experimental data from hadron colliders and lepton-hadron scattering, the momentum distributions of quarks and gluons (parton distribution functions, PDFs) are now known to per-cent level accuracy~\cite{Gao:2017yyd,Bailey:2020ooq,Hou:2019efy,NNPDF:2021njg}. 

The understanding 
of other aspects of the 
proton structure is much less well-developed. Most notably, information on the proton’s spin 
structure~\cite{Aidala:2012mv} is 
still quite sparse, relying on a limited set of data from polarized lepton-nucleon scattering at fixed target
energies and from polarized proton-proton collisions at the BNL RHIC collider. It is one of the primary 
objectives of the planned electron-ion collider (EIC) at BNL to provide in-depth probes of the nucleon 
spin structure through a variety of different measurements in polarized electron-proton collisions~\cite{Accardi:2012qut}. 

An essential aspect of the proton’s spin structure is encoded in polarization-dependent PDFs,  
describing the probability of finding a parton with a given momentum fraction $x$ at a resolution 
scale $Q^2$ with its helicity aligned or anti-aligned to the nucleon’s spin. One considers the 
unpolarized and polarized combinations: 
\begin{align}
 f(x,Q^2) =& f^+(x,Q^2)+f^-(x,Q^2)\,, \nonumber\\
 \Delta f(x,Q^2)=&  f^+(x,Q^2)-f^-(x,Q^2)\,,
\end{align}
where $^\pm$ refers to the relative orientation of the parton helicity with respect to the parent nucleon spin.
$ f(x,Q^2)$ are the well-established (unpolarized) PDFs, 
while $\Delta f(x,Q^2)$ contain the essential information on the spin structure of the nucleon. 
Experimental probes of the polarized PDFs $\Delta f(x,Q^2)$ rely on the measurement of asymmetries 
 in collisions where both probe and target are longitudinally polarized. 

The main experimental information entering into the 
determination of polarized 
PDFs~\cite{Nocera:2014gqa,deFlorian:2014yva}
comes from polarized deeply 
inelastic lepton-nucleon 
scattering~\cite{EuropeanMuon:1989yki,SpinMuon:1998eqa,E142:1996thl,E143:1998hbs,E154:1997xfa,E155:2000qdr,HERMES:2006jyl,COMPASS:2015mhb}, which probes specific 
charge-weighted combinations of the polarized quark distributions. 
Semi-inclusive identified hadron production in deep inelastic scattering (SIDIS)
offers valuable supplementary 
information~\cite{SpinMuon:1997yns,COMPASS:2010hwr,HERMES:2018awh},
since the identified hadron species can be associated 
to the flavor of the parton that was produced in the underlying
hard scattering process. 
 This connection is described by fragmentation functions (FFs), encoding the probability of a parton fragmenting into a hadron of a given type. Similar to the PDFs, FFs are universal, non-perturbative objects that factorize from 
 the parton-level subprocess and whose evolution with the 
 resolution scale is 
 determined by the DGLAP 
 equations.
Especially future SIDIS measurements at the BNL EIC 
will provide an indispensable tool to 
disentangle the precise flavor structure of the 
nucleon spin~\cite{AbdulKhalek:2021gbh,Aschenauer:2019kzf}.

Global fits of the unpolarized PDFs are routinely performed 
at next-to-leading order (NLO) and next-to-next-to-leading order (NNLO) in QCD, thereby requiring corrections at the  appropriate order to the evolution kernels and to the parton-level cross sections
for all experimental observables that 
are included in the fits.

For polarized PDFs, the evolution kernels are known to 
NLO~\cite{Vogelsang:1996im,Mertig:1995ny} and 
NNLO~\cite{Moch:2014sna,Blumlein:2021enk,Blumlein:2021ryt}, but  corrections
to parton-level coefficient functions beyond NLO have up to now been 
obtained
only for inclusive polarized 
DIS~\cite{Zijlstra:1993sh,Blumlein:2022gpp}. The polarized SIDIS coefficient functions are currently known 
to NLO~\cite{deFlorian:1997zj}.
Following up on initial leading-order (LO)
studies~\cite{Gluck:1988uj,Altarelli:1988mu}, 
polarized PDFs have been determined routinely at NLO through 
global fits~\cite{Gluck:1995yr,Gehrmann:1995ag,AsymmetryAnalysis:1999gsr,deFlorian:2008mr,deFlorian:2009vb,deFlorian:2014yva,Blumlein:2010rn,Nocera:2014gqa}
to spin-asymmetry data.

In view of future precision 
data from the BNL EIC, an extension of 
polarized PDF studies to NNLO accuracy would be very much 
desirable.  
With the enhanced precision and including 
polarized SIDIS measurements, these 
will not only help disentangle the PDFs of polarized gluons, valence and sea quarks, but also allow to discern quark and anti-quark contributions to the polarized flavor PDFs.
First steps in this direction were taken most 
recently, with NNLO fits by two groups~\cite{Bertone:2024taw,Borsa:2024mss} to
polarized inclusive DIS and SIDIS data, approximating~\cite{Abele:2021nyo} the 
NNLO SIDIS coefficient functions from a threshold expansion. 

With this letter, we enable consistent
precision studies by computing
the NNLO QCD coefficient functions for longitudinally 
polarized SIDIS, following up on recent NNLO results obtained 
for unpolarized SIDIS~\cite{Goyal:2023zdi,Bonino:2024qbh}.

\section{Kinematics of polarized SIDIS}

We consider the observation of an unpolarized hadron $h$ from the scattering of a polarized lepton off a polarized nucleon. 
Both polarizations are longitudinal. Following the notation of~\cite{deFlorian:1997zj,Anderle:2013lka}, we describe polarized semi-inclusive deep-inelastic scattering as
$\vec{\ell}(k)\,\vec{p}(P)\to\ell(k^{\prime})\,h(P_h)\,X$, with some inclusive final-state
radiation $X$. The vector $q=k-k^{\prime}$ denotes the momentum transfer
between the leptonic and hadronic systems, and $y=(P\cdot
q)/(P\cdot k)$ the associated energy transfer at 
virtuality $Q^2=-q^2$. The quantities
\begin{align}
x=&\frac{Q^2}{2P\cdot q} & \rm{and}& &  z=&\frac{P\cdot P_h}{P\cdot q}
\end{align}
are the momentum fractions of the nucleon carried by the incoming parton
($x$), and of the outgoing parton carried by the identified
hadron ($z$) at Born level. The center-of-mass energy of the lepton-nucleon system  $\sqrt{s}$
is given by $s=Q^2/(xy)$.

The experimentally measured double spin asymmetry is obtained  
from the difference of anti-aligned and aligned spin orientations of probe and target~\cite{HERMES:2004zsh}. After correcting for QED effects and neglecting higher-twist contributions, it can be expressed as ratio of 
hadronic SIDIS structure functions as 
\begin{align}\label{eq:A1h}
A^h_1(x,z,Q^2)
= \frac{g^h_1(x,z,Q^2)}{F^h_1(x,z,Q^2)} \, .
\end{align}
The $g^h_1$ polarized SIDIS structure function 
receives contributions from different partonic channels. 
 These are given by the convolution
between the polarized PDF $\Delta f_p$ for a parton $p$, the FF $D^h_{p^{\prime}}$ of parton $p^{\prime}$ into hadron $h$, and the polarized coefficient function $\Delta \Ccal_{p' p}$ for the partonic
transition $p\to p^{\prime}$:
\begin{align}\label{eq:g1h}
  2g^h_1&(x,z,Q^2)= \sum_{p,p'} \int_x^1 \frac{\der\hx}{\hx}
  \int_z^1 \frac{\der\hz}{\hz} \Delta f_p\left(\frac{x}{\hx},\mu_F^2\right)
   \nonumber \\ 
  &\times\, D_{p'}^h\left(\frac{z}{\hz},\mu_A^2\right)  
 \Delta \Ccal_{p' p}\left(\hx,\hz,Q^2,\mu_R^2,\mu_F^2,\mu_A^2\right).
\end{align}
Likewise, the unpolarized SIDIS structure function $F_1^h$ reads
\begin{align}\label{eq:F1h}
  2F^h_1&(x,z,Q^2)= \sum_{p,p'} \int_x^1 \frac{\der\hx}{\hx}
  \int_z^1 \frac{\der\hz}{\hz} f_p\left(\frac{x}{\hx},\mu_F^2\right)
   \nonumber \\ 
  &\times\, D_{p'}^h\left(\frac{z}{\hz},\mu_A^2\right)  
  \Ccal^T_{p' p}\left(\hx,\hz,Q^2,\mu_R^2,\mu_F^2,\mu_A^2\right) . 
\end{align}

In the above expressions, $\mu_F$  and $\mu_A$
denote the mass factorization scales of PDFs and FFs, 
while $\mu_R$ is the renormalization scale. 
 The SIDIS coefficient
functions $(\Delta)C_{p'p}$ encode the hard-scattering part of the process, and can be computed in
perturbative QCD. Their perturbative expansion in the strong coupling constant
$\alpha_s$ reads
\begin{align}
  (\Delta) \Ccal_{p' p} =& (\Delta)C^{(0)}_{p' p}
  + \frac{\alpha_s(\mu_R^2)}{2\pi}  (\Delta)C^{(1)}_{p' p}
  \nonumber \\ 
  &+ \left(\frac{\alpha_s(\mu_R^2)}{2\pi}\right)^2   (\Delta)C^{(2)}_{p' p}
  + \mathcal{O}(\alpha_s^3)\, .
\end{align} 

At LO, only the $qq$ channel ($\gamma^* q\to q$) contributes, with the LO
coefficient function normalized to
\begin{equation}
  \DC^{(0)}_{qq}= e_q^2 \delta(1-\hx) \delta(1-\hz)\, ,
\end{equation}
where $e_q$ is the quark charge. 
At NLO~\cite{deFlorian:1997zj}, the channels $qg$ and
$gq$ start to contribute, yielding $\DC^{(1)}_{qq}$, $\DC^{(1)}_{gq}$ and $\DC^{(1)}_{qg}$.

In this letter we present results for the NNLO corrections
$\DC^{(2)}_{p^{\prime}p}$ to all partonic channels appearing at this order.
Following the notation of~\cite{Anderle:2016kwa,Bonino:2024qbh}, the seven partonic channels
appearing at $\mathcal{O}(\alpha_s^2)$ are:
{ \allowdisplaybreaks
\begin{align}\label{CFNNLOlist}
\DC^{(2)}_{qq}&=e_q^2 \DC^{\mathrm{NS}}_{qq}+\biggl( \sum_j e^2_{q_j}\biggr)\DC^{\mathrm{PS}}_{qq} \, , \nonumber \\
\DC^{(2)}_{\qb q}&=e_q^2\DC_{\qb q} \, , \nonumber \\
\DC^{(2)}_{\qp q}&=e_q^2 \DC^{1}_{\qp q}+e_{\qp}^2 \DC^{2}_{\qp q}+e_q e_{\qp}\DC^{3}_{\qp q} \, , \nonumber \\
\DC^{(2)}_{\qbp q}&=e_q^2 \DC^{1}_{\qp q}+e_{\qp}^2 \DC^{2}_{\qp q}-e_q e_{\qp} \DC^{3}_{\qp q} \, , \nonumber \\
\DC^{(2)}_{gq}&=e_q^2 \DC_{gq} \, , \nonumber \\
\DC^{(2)}_{qg}&=e_q^2 \DC_{qg} \, , \nonumber \\
\DC^{(2)}_{gg}&=\biggl( \sum_j e_{q_j}^2 \biggr) \DC_{gg} \,  .
\end{align}
}
With $\overset{\textbf{\fontsize{5pt}{5pt}\selectfont(--)}\prime}{q\phantom{.}}$ we indicate an (anti-)quark of flavor different from $q$.

\section{Analytical calculation}
To obtain the matrix elements relevant to the polarized coefficient functions, we use the projectors from the inclusive calculation~\cite{Zijlstra:1993sh}. 
The appearance of the inherently four-dimensional objects $\gamma_5$ and $\varepsilon^{\mu\nu\rho\sigma}$ in the external projectors requires a consistent treatment in dimensional regularization \cite{tHooft:1972tcz}. 
A common choice is the 't Hooft-Veltman-Breitenlohner-Maison (\HVBM) scheme \cite{tHooft:1972tcz,Breitenlohner:1977hr}. 
The Larin prescription \cite{Larin:1991tj,Larin:1993tq} is derived from the \HVBM{} scheme, and consists of setting
\begin{equation}
    \gamma_\mu \gamma_5 = \frac{i}{3!} \varepsilon_{\mu\nu\rho\sigma}\gamma^{\nu}\gamma^{\rho}\gamma^{\sigma}
\end{equation}
and evaluating the Dirac traces in $d$ dimensions. 
The two remaining Levi-Civita tensors are contracted into $d$-dimensional metric tensors.
Quantities in the Larin scheme will be denoted by an upper index $\Lar$, and are subsequently converted to the $\MSbar$ \nolinebreak scheme.

The NNLO corrections consist of three types of contributions, defined relative to the underlying  Born process:
 double-real (RR) corrections stem from two additional radiations, the real-virtual (RV) corrections from a one-loop correction and an additional radiation, and the double-virtual (VV) corrections from two-loop virtual insertions.

The evaluation and integration of the RR and RV 
contributions 
proceed like in the unpolarized calculation \cite{Bonino:2024qbh}: following integration-by-part (IBP) reduction \cite{Chetyrkin:1981qh,Laporta:2000dsw} with \texttt{Reduze2} \cite{vonManteuffel:2012np}, the RR contributions are 
expressed in terms of 21 master integrals, which were 
computed by solving their differential equations~\cite{Gehrmann:1999as}. 
Their analytic
expressions 
are presented in \cite{Bonino:2024adk}. The RV integrals are solved by analytically integrating out the loop~\cite{Gehrmann:1999as}, then
continuing the appearing hypergeometric functions to the 
appropriate  Riemann sheets  and subsequently
expanding out the phase space measure in terms of 
distributions in each of the sheets as 
 described in \cite{Gehrmann:2022cih}. The master integrals for the VV part are taken from \cite{Gehrmann:2005pd}.

The calculation is carried out using \texttt{Mathematica} and \texttt{FORM} \cite{Vermaseren:2000nd}. We express the integrals in terms of harmonic polylogarithms (HPLs) \cite{Remiddi:1999ew} with the help of the packages \texttt{HPL} \cite{Maitre:2005uu} and \texttt{PolyLogTools} \cite{Duhr:2019tlz}.

Interestingly, the NLO virtual (V) 
and the NNLO VV contributions to $g_1^h$ 
are described by the respective vector form factors \cite{Gehrmann:2005pd} rather than by their axial counterparts: 
the photon couples to the quark line through a vector coupling, whereas the antisymmetric current carried by the photon is contracted only from the external leg. As a consequence, traces of quark-loops coupling to the polarized photon can be carried out consistently in $d=4-2\epsilon$ dimensions without giving rise to the axial anomaly \cite{Adler:1969gk,Bell:1969ts}.
This stands in sharp contrast to calculations of operator matrix elements (OME), e.g.~\cite{Matiounine:1998re}, where the $g_1$ projector of the photon is absorbed into an operator insertion, rendering the photon coupling axial, and giving rise to the anomaly.

We renormalize $\alpha_s$ in the $\MSbar$ scheme to remove poles of 
ultraviolet origin. The remaining infrared poles can be eliminated by mass factorization 
on the polarized PDFs and unpolarized FFs. 
At this stage, all coefficient functions are formulated in the Larin scheme. Consequently, the polarized PDF mass factorization 
counterterms are taken
in the Larin scheme, constructed from the 
polarized spacelike splitting functions in this scheme. 

These splitting functions \cite{Mertig:1995ny,Vogelsang:1996im} in Larin and $\MSbar$ are related by a scheme transformation \cite{Zijlstra:1993sh}, which follows from the 
scheme 
invariance of the measurable inclusive structure function
$g_1$~\cite{Matiounine:1998re}:
\begin{align}
    g_1 =& \Delta \Ccal^{\MSbar} \otimes \Delta f^{\MSbar} \nonumber \\
    =& (\Delta \Ccal^{\Lar} \otimes Z^{-1}) \otimes ( Z \otimes \Delta f^{\Lar})
    = \Delta \Ccal^{\Lar} \otimes \Delta f^{\Lar}
\label{eq:scheme}
\end{align}
with 
\begin{align}
    \Delta \Ccal=& \begin{pmatrix}
        \Delta \Ccal_q\\
        \Delta \Ccal_g
    \end{pmatrix}^T, &
    \Delta f =& \begin{pmatrix}
        \Delta q \\
        \Delta g
    \end{pmatrix}, &
    Z =& \begin{pmatrix}
        z_{qq} & z_{qg} \\
        z_{gq} & z_{gg}
    \end{pmatrix},
\end{align}
where the entries of $Z$ in non-singlet and singlet case
are given in \cite{Moch:2014sna}. All convolutions are carried out by using \texttt{MT} \cite{Hoschele:2013pvt}.

In order to obtain the $\MSbar$ SIDIS coefficient functions, we scheme-transform the Larin-scheme coefficient functions
according to (\ref{eq:scheme}).  
Our results with full scale dependence ($\mu_R$, $\mu_F$, $\mu_A$) are provided in an ancillary file to the \texttt{arXiv} submission of this paper.

Several checks were performed to validate our results. 
We verified that our results in each of the channels in~\eqref{CFNNLOlist}
fulfill the renormalization group equations in both Larin and $\MSbar$ scheme.
Moreover, we used the underlying RR, RV and VV 
polarized subprocess 
matrix elements to re-derive the inclusive NNLO coefficient 
function for $g_1$, finding 
full agreement with~\cite{Zijlstra:1993sh}. We subsequently 
integrated specific subprocess contributions to the 
SIDIS coefficient functions over the 
final-state parton momentum fraction $z$, recovering 
the respective  contributions to the $g_1$ coefficient 
functions. 
We also compared our leading and subleading power terms for the $\Delta C_{qq}^{(2)}$ coefficient function against the prediction from NNLL threshold-resummation in \cite{Abele:2021nyo},  finding full agreement.
Finally, an independent calculation of the polarized SIDIS coefficient functions at NNLO was performed by another group~\cite{Goyal:2024tmo}
 in close timely coincidence with our results, finding mutual agreement in all channels.   
\begin{figure*}[t]
  \includegraphics[width=0.78\textwidth]{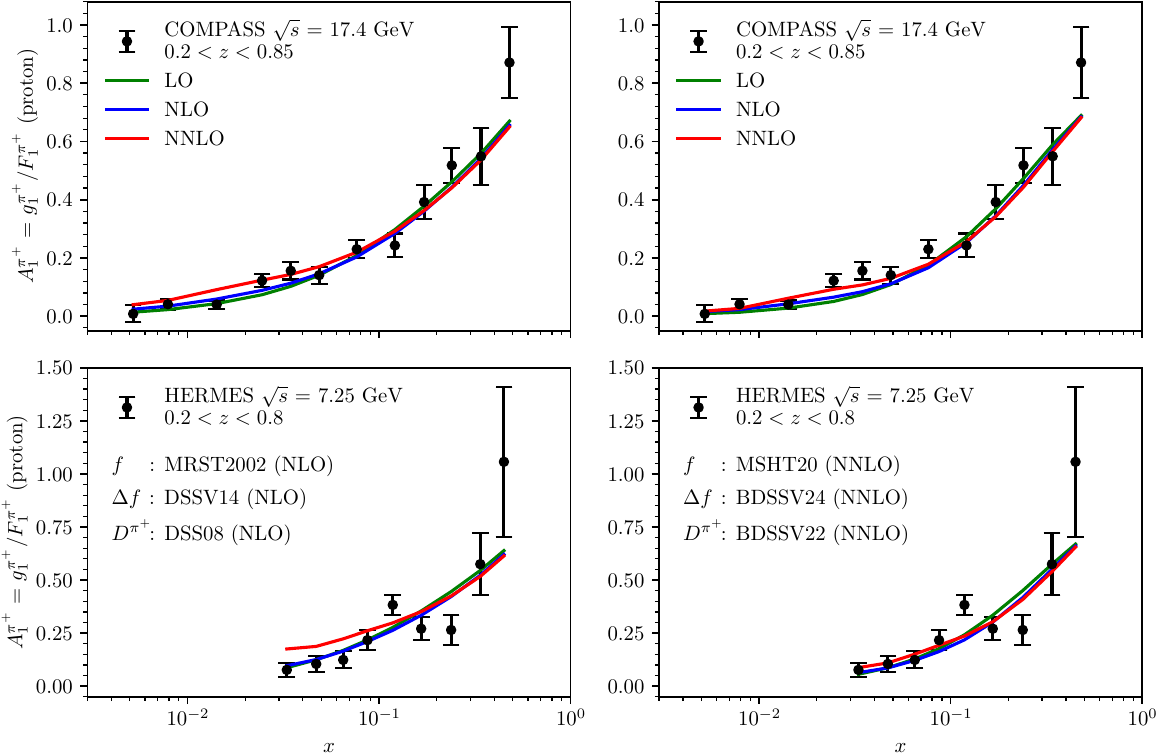}
  \caption{Asymmetries $A_1^{\pi^+}$ at different perturbative orders compared to data 
  from COMPASS~\protect\cite{COMPASS:2010hwr}  
  and HERMES~\protect\cite{HERMES:2018awh}, 
  computed using 
  NLO PDFs and FFs (left) and NNLO PDFs and FFs (right).}
  \label{fig:compassandhermesdata}
\end{figure*}

\section{Results}

Measurements of SIDIS for various species of final state hadrons
on proton targets were performed by 
CERN COMPASS \cite{COMPASS:2010hwr} and the DESY HERMES \cite{HERMES:2018awh} fixed-target 
experiments. The experiments present their results in 
terms of the longitudinal double spin asymmetry $A_1^h$~\eqref{eq:A1h}. COMPASS~\cite{COMPASS:2010hwr} 
measured 
at $\sqrt{s}=17.4$~GeV
and included all hadrons produced in the 
range $0.2<z<0.85$. The HERMES~\cite{HERMES:2018awh} 
data were taken 
at $\sqrt{s}=7.25$~GeV and included hadrons in the range $0.2<z<0.85$.
 The asymmetry $A_1^h$ is evaluated at central bin values in $x$ and $Q^2$, justified by a relatively narrow range in $Q^2$ for each bin due to the kinematical restrictions of the experiments, at $\mu_R=\mu_F=\mu_A=\sqrt{Q^2}$.

To illustrate the 
numerical impact of the newly computed NNLO corrections to the 
polarized SIDIS coefficient functions, we compare COMPASS data~\cite{COMPASS:2010hwr}  for $A_1^{\pi^+}$ 
to predictions at  LO, NLO and NNLO. To single out the impact of the coefficient functions, 
predictions at the different orders are computed with the same 
sets of parton distributions and fragmentation functions. 
We display predictions 
for two different setups: 
(A) PDFs and FFs at NLO throughout: we use 
the polarized NLO PDF DSSV14 set~\cite{deFlorian:2014yva},
in combination (as used in the original 
DSSV14 fit) with unpolarized NLO PDFs from the 
MRST2002 set~\cite{Martin:2002aw} and NLO FFs from the 
DSS08 set~\cite{deFlorian:2007aj}.  For $\alpha_s$, we use the \texttt{APFEL++}~\cite{Bertone:2013vaa,Bertone:2017gds} routine at NLO with reference value $\Lambda^{(n_f=4)}_{\mathrm{QCD}}=334\, \MeV$, corresponding to $\alpha_s(M_Z)=0.119$. (B) PDFs and FFs at NNLO 
throughout: we use the polarized NNLO BDSSV24
set~\cite{Borsa:2024mss}, with unpolarized NNLO PDFs from 
MSHT20~\cite{Bailey:2020ooq} and NNLO FFs from
BDSSV22~\cite{Borsa:2022vvp} and $\alpha_s$ at NNLO through
 LHAPDF~\cite{Buckley:2014ana},  with $\alpha_s(M_Z)=0.118$.

Figure~\ref{fig:compassandhermesdata} compares 
the predictions for 
$A_1^{\pi^+}$
from COMPASS~\protect\cite{COMPASS:2010hwr}  
  and HERMES~\protect\cite{HERMES:2018awh}. We 
  observe that the predictions for the asymmetry are 
  very stable between LO and NLO. Including  
  the NNLO corrections slightly decreases the 
  asymmetry at high $x$ and leads to a small enhancement 
  of the asymmetry at lower values of $x$. These changes 
  are more pronounced for the predictions using NLO PDFs and FFs (left frames). Both the COMPASS  and HERMES
  datasets on $A_1^{\pi^+}$ were included in the 
  DSSV14 NLO and BDSSV24 NNLO fits of polarized PDFs, using 
  exact SIDIS coefficient functions at NLO and an 
  approximation~\cite{Abele:2021nyo} at NNLO. 
The numerical magnitude of the NNLO corrections to the 
SIDIS asymmetry $A_1^{\pi^+}$ at low $x$ clearly demonstrates 
the potential impact of the exact corrections and calls 
for a careful reassessment of the impact of SIDIS data 
in a future global fit at NNLO. 
\begin{figure}[t]
  \includegraphics[width=0.47\textwidth]{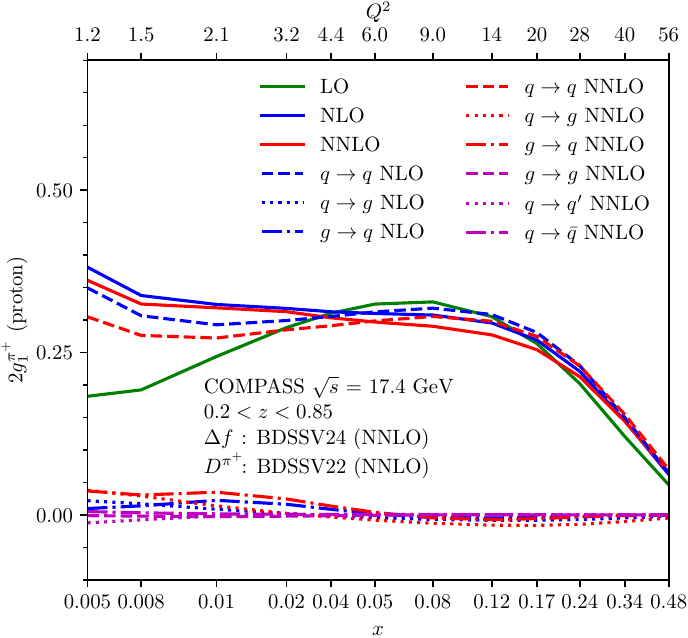}
  \caption{Channel decomposition of theory predictions for $g_1^{\pi^+}$ with COMPASS kinematics.}
  \label{fig:compassg1h}
\end{figure}

In Figure~\ref{fig:compassg1h} we study the impact of each partonic channel on $g_1^{\pi^+}$ at LO, NLO and NNLO for COMPASS kinematics. We use the NNLO BDSSV24
PDFs~\cite{Borsa:2024mss} and BDSSV22 FFs~\cite{Borsa:2022vvp} throughout.
For $x$ and $Q^2$ we use the central values of the kinematical bins from the COMPASS analysis~\cite{COMPASS:2010hwr}, integrating over $z\in[0.2,0.85]$.
While the $q\to q$ channel remains the dominant contribution, the $q\to g$ channel plays a more prominent role at NNLO than at NLO, but in the small-$x$ region also the $g \to q$ channel provides a sizable contribution starting from NNLO.
We notice a reduction in the size of the NNLO corrections for larger values of $x$ (and in turn of $Q^2$), which points at an improvement in the convergence of the perturbative series with increasing $Q^2$. 

\begin{figure}[t]
    \centering
    \includegraphics[width=0.47\textwidth]{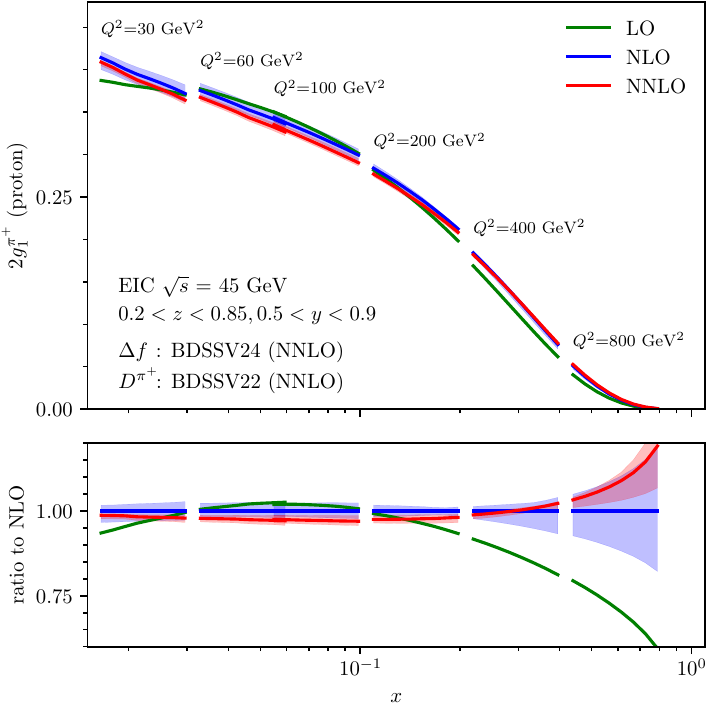}
    \caption{Longitudinally polarized SIDIS structure 
    function $g_1^{\pi^+}$ for EIC kinematics at $\sqrt{s}=45$~GeV, evaluated at different perturbative orders.}
    \label{fig:eic45g1}
\end{figure}

To  study the 
perturbative convergence of the polarized SIDIS
predictions
at higher energies, Fig.~\ref{fig:eic45g1} displays 
$g_1^{\pi^+}$ (integrated over $0.2<z<0.85$) of the proton 
for EIC-like kinematics with $\sqrt{s}=45$~GeV. 
We consider several values of $Q^2$, and for each value of $Q^2$ we plot predictions for values of $x$ constrained by the requirement $0.5 < y < 0.9$.
Using the 
 polarized PDFs and FFs at NNLO throughout allows 
us to assess the uncertainty on the theory 
predictions in a consistent manner 
through a seven-point scale variation on $\mu_R$ and $\mu_F=\mu_A$ around their central value
 $\sqrt{Q^2}$, discarding opposite variations of any pair of scales.

We observe very good perturbative convergence of the predictions. While the NLO contributions enhance  
the LO prediction by 5--30\%, we note that in the range $0.03 < x < 0.5$, inclusion of the 
 NNLO contribution changes the predictions by less than 
  $\pm 5\%$, slightly increasing towards smaller or larger values of $x$. 
  The scale uncertainty at NLO ranges between $\pm 5\%$ at low $x$ values to $\pm 15\%$
at high $x$. At NNLO, it is reduced throughout the kinematical range to $\pm 2\%$ at low $(x,Q^2)$ and 
at most $\pm 8\%$ at larger values. The non-trivial kinematical shape of the NNLO corrections 
highlights their importance for precision physics studies at the EIC.

\section{Conclusions}
Important information on the quark flavor decomposition of
the proton spin content is gained from polarization 
asymmetries in SIDIS. We computed the NNLO QCD corrections 
to the polarized SIDIS coefficient functions, which 
turn out to have a sizable numerical impact 
on the asymmetries especially at low 
$x$ and low $Q^2$. Our results allow to include
polarized SIDIS data 
in future  NNLO precision studies of polarized parton 
distributions, thereby enabling an unprecedented
level of detail in the understanding of the 
spin structure of the proton.

\begin{acknowledgments}
\sect{Acknowledgments}
We would like to thank the authors of~\cite{Goyal:2024tmo} for 
discussions on the comparison of results. 
We are grateful to Daniel de Florian for providing us with the DSSV PDF set~\cite{deFlorian:2014yva}, and to Ignacio Borsa for the BDSSV PDF set~\cite{Borsa:2024mss} and detailed validations of the numerical 
implementation of the NNLO coefficient functions.
This work has received funding from the Swiss National Science Foundation (SNF)
under contract 200020-204200 and from the European Research Council (ERC) under
the European Union's Horizon 2020 research and innovation program grant
agreement 101019620 (ERC Advanced Grant TOPUP).
\end{acknowledgments}

\bibliography{polsidis}

\end{document}